\documentclass{epl}

\title{Layer dependent band dispersion and correlations using tunable Soft 
X-ray ARPES}
\author{N. Kamakura\inst{1}, Y. Takata\inst{1}, T. Tokushima\inst{1},
Y. Harada\inst{1}, A. Chainani\inst{1,2}, K. Kobayashi\inst{3} \and 
S. Shin\inst{1,4}}
\institute{
  \inst{1} RIKEN, Harima Institute, 1-1-1 Kouto, Mikazuki, Sayo,
Hyogo 679-5148, Japan\\
\inst{2} Institute for Plasma Research, Bhat, Gandhinagar 382 428, Gujarat, 
India\\
\inst{3} SPring-8/JASRI Mikazuki, Hyogo 679-5198, Japan\\
\inst{4} Institute for Solid State Physics, University of Tokyo, 
  Kashiwanoha, Kashiwa, Chiba 277-8581, Japan
}


\begin{document}

\maketitle

\begin{abstract}
Soft X-ray Angle-Resolved Photoemission Spectroscopy is applied to study 
in-plane band dispersions of Nickel as a function of probing depth. Photon 
energies between $h\nu$ = 190 and 780 eV were used to effectively probe up to 
$\sim$3--7 layers ($\sim$5--12 \AA). The results show layer 
dependent band dispersion of the $\Delta_{2\downarrow}$ minority-spin
band which crosses 
the Fermi level in 3 or more layers, in contrast to known top 1--2 layers 
dispersion obtained using ultra-violet rays. The layer dependence 
corresponds to an increased value of exchange splitting and 
suggests reduced 
correlation effects in the bulk compared to the surface.
\end{abstract}

\section{Introduction}
Angle-Resolved Photoemission Spectroscopy (ARPES) is a valuable tool to 
study the experimental 
band-structure(BS), $w(\mathbf{k})$, where $w$ is energy and $\mathbf{k}$ 
is momentum
of electrons in a solid. Recent Ultra-violet-ARPES 
(ARUPS) studies of correlated materials have
provided important results like spin and charge collective 
modes in a quasi 1-D (dimensional) metal~\cite{Ref01}, 
dimensional crossover~\cite{Ref02}, exotic 
characteristics of the high-$T_{c}$ 
cuprates~\cite{Ref03}, etc.\ While ARUPS is extremely well suited to 
study electronic 
BS of low-dimensional solids, it probes the top 1--2 layers of the 
surface~\cite{Ref04}. 
In order to determine bulk BS of 3-D correlated systems which can 
depend on the probed layer~\cite{Ref05,Ref06,Ref07,Ref08}, it is meaningful 
to use higher energies. 
As the excitation energy is increased, the probing depth or mean free path(MFP) increases~\cite{Ref09}, but to date, no 
layer dependent variation of in-plane band dispersion (BD) 
beyond the top 2 layers has been 
reported. In this work, using tunable soft X-ray ARPES, 
we study layer dependence of in-plane BDs of Nickel metal.

Nickel is a prototype of a correlated ferromagnetic metal and has been 
extensively studied using PES and inverse-PES spectroscopies
~\cite{Ref10,Ref11,Ref12,Ref13,Ref14,Ref15,Ref16,Ref17,Ref18,Ref19}. 
Beginning with the work of 
Slater~\cite{Ref20} and Stoner~\cite{Ref21}, the BS of Nickel has remained a 
suitable testing ground for a variety of experiments and theory. Recent 
theories have addressed the major inconsistencies with experiment : 
3$d$ bandwidth extent,  temperature(T)-dependence of magnetization, 
and a satellite structure which appears at about 6 eV below the Fermi 
level($E_{F}$). The 3$d$ bandwidth and exchange splitting(ES) of 
Nickel observed by ARUPS are reduced 
by 25\% and 50\%, respectively~\cite{Ref10,Ref11,Ref12,Ref13,Ref14,Ref15}, 
compared to the values obtained from 
spin-polarized local density approximation (s-LDA) 
calculations~\cite{Ref22}. The 6 eV satellite obtained in 
PES experiments~\cite{Ref16,Ref17,Ref18} is 
the two-hole bound state due to correlation effects. It 
is missing in BS calculations in the one-electron 
picture~\cite{Ref22} but obtained in ``generalized Hubbard 
models"~\cite{Ref14,Ref19,Ref23} or the ``LDA (GWA) + DMFT" 
(Dynamical Mean-Field Theory)~\cite{Ref24,Ref25} which explicitly 
include Coulomb interactions. A recent study discusses 
the inability of s-LDA to reproduce the position of the 
minority spin state $X_{2\downarrow}$ below $E_{F}$ obtained by 
ARUPS~\cite{Ref14}. 
The energy position of $X_{2\downarrow}$ 
states determines the electron count associated with the hole pocket of 
minority spin character at $X$-point, which in turn influences the 
magnetic 
moment, a macroscopic property. Although recent theoretical efforts 
reproduce the experimental BS of Nickel 
by including correlation effects~\cite{Ref14,Ref25}, the results were 
compared with experimental BS obtained using ARUPS. The present study 
reports new experimental results of the bulk BS of Nickel obtained using 
soft X-ray ARPES.

\section{Experiment}
Experiments were performed at beam line 27SU of SPring-8~\cite{Ref26} using 
linearly polarized light. Total energy resolution was 50--160 meV. 
The beam line has a figure-8 undulator~\cite{Ref27}, enabling an easy 
switch of the polarization vector from horizontal(H)- to vertical(V)-
polarization. Ni(100) surface was prepared by Ar$^+$ sputtering and 
annealing. The surface was checked by core level photoemission spectra 
obtained using 780 eV photons and the contamination due to Oxygen and 
Carbon was measured to be $< 1\%$. The surface crystallinity was confirmed 
to be a sharp (1$\times$1) LEED pattern.

\begin{figure}[b]
\onefigure[width=140mm]{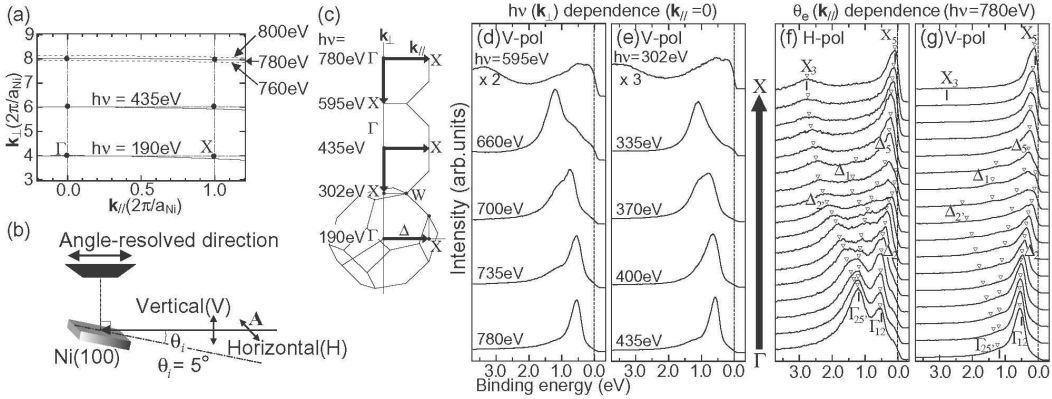}
\caption{(a) $\mathbf{k}_{\bot}$ versus $\mathbf{k}_{\|}$ for soft X-ray 
ARPES of Ni(100) for $h\nu$ = 190, 435 and 780 eV photons by the 
experimental geometry used(fig.\ 1(b)). Fig.\ 1(c) shows the 3-D BZ 
of Ni(100) alongwith the probed sections. For $\mathbf{k}_{\|}$ = 0 
and $h\nu$ = (d) 780--595 eV and (e) 435--302 eV, nearly identical 
spectra in successive BZs, which are normalized by photon flux, 
verify the validity of the $\mathbf{k}_{\bot}$ points shown in (a) and (c). 
Soft X-ray ($h\nu$ = 780 eV) ARPES of Ni(100) along $\Gamma$-$X$ 
($\Delta$-line) with (f) H and (g) V-polarization, where each spectrum 
is integrated over $0.5^{\circ}$.  Peak positions ($\bigtriangledown$) 
indicate BDs ($\Delta_{2}$, $\Delta_{5}$, $\Delta_{1}$ and $\Delta_{2'}$).}
\label{f.1}
\end{figure}

ARPES measures $w$ and $\mathbf{k}$ of electrons in a solid according to the 
following equations~\cite{Ref04}.

\begin{equation}
\label{e.1}
\hbar \mathbf{k}_{\|} = \sqrt{2m(h\nu - w - \phi )}\sin \theta_{e}\ ,
\end{equation}

\begin{equation}
\label{e.2}
\hbar \mathbf{k}_{\bot} = \sqrt{2m\{(h\nu - w - \phi )\cos^2 \theta_{e} 
+ V_{0}\}}\ ,
\end{equation}

where $\mathbf{k}_{\|}$ and $\mathbf{k}_{\bot}$ are parallel and 
perpendicular components of $\mathbf{k}$, respectively, $\theta_{e}$ 
is electron emission angle, $\phi$ is
work function, and $V_{0}$ is inner potential. 
The angular resolution of the electron energy analyzer 
was better than $\pm$0.2$^\circ$ which corresponds to resolutions of about 
$\pm$0.014, $\pm$0.021, and $\pm$0.028 ($2\pi/a_{Ni}$) in $\mathbf{k}_{\|}$ 
at $h\nu$ = 190, 435, and 780 eV, respectively (eq.~(\ref{e.1})). 
Eqs.~(\ref{e.1}) and ~(\ref{e.2}) indicate that 
when ARPES with fixed $h\nu$ is used for measurement of in-plane 
($\mathbf{k}_{\|}$ ) BDs, 
$\mathbf{k}_{\bot}$ also changes. The variation of 
$\mathbf{k}_{\bot}$ can be substantial in ARUPS~\cite{Ref28}, 
but diminishes with increasing $h\nu$, making it an advantage. 
In fig.~\ref{f.1}(a), we plot $\mathbf{k}_{\bot}$ 
versus $\mathbf{k}_{\|}$ in units of ($2\pi/a_{Ni}$) 
for $h\nu$ = 190, 435,\ and 780\ eV in the experimental 
geometry of fig.~\ref{f.1}(b), using eqs.~(\ref{e.1}) and ~(\ref{e.2}) 
with $V_{0}$ = 9.3 eV. Actually, the momentum 
transfer of the photon is not negligible in soft X-ray ARPES. 
However, since our experimental geometry is near grazing 
incidence (fig.~\ref{f.1}(b)), the momentum transfer of the photon 
results in a constant shift of parallel component of the 
momentum $\mathbf{k}_{\|}$, and $\mathbf{k}_{\bot}$ is 
negligibly influenced~\cite{Ref29}. Therefore, from fig.~\ref{f.1}(a) 
the spectra of $\mathbf{k}_{\|}$ = 0 at 
these $h\nu$ (190, 435, and 780 eV) are expected to probe an equivalent 
$\mathbf{k}_{\bot}$ point 
($\Gamma$) in the 3-D Brillouin zone (BZ) of Ni(100)(fig.~\ref{f.1}(c)). 
We first validate this by $h\nu$-dependent ARPES of 
$\mathbf{k}_{\|}$ = 0 in successive 
BZs (figs.~\ref{f.1}(d) and ~\ref{f.1}(e)). (The spectra of 
$\mathbf{k}_{\|}$ = 0 are identified from the observed band dispersion 
in the ($\theta_{e}$-dependent) ARPES at each photon energy.) 
The changes in peak positions 
correspond to $\mathbf{k}_{\bot}$ BD, 
mainly due to $\Delta_{1}$ band. The similarity 
of spectra in figs.~\ref{f.1}(d) and ~\ref{f.1}(e) measured in successive BZs 
confirm we are at equivalent momentum regions and the $\Gamma$-point 
can be precisely measured 
according to eq.~(\ref{e.2}) even by soft X-ray ARPES. 
Further, since the difference of $\mathbf{k}_{\bot}$ 
along $\Gamma$-$X$ ($\Delta$-line)of $\mathbf{k}_{\|}$ is 
already small with 190 eV photons 
(fig.~\ref{f.1}(a)), the in-plane BD along $\Gamma$-$X$ can be measured by 
ARPES with 190, 435, and 780 eV photons. The probing depth or MFP is 
estimated to be about 5, 8, and 12 \AA\ at $h\nu$ = 190, 435, and 780 eV, 
respectively~\cite{Ref09}. Thus we can probe up to the third, fifth, 
and seventh layer of equivalent regions in the 3-D 
BZ (figs.~\ref{f.1}(a)-(c)).

ARPES with soft X-rays can also lead to indirect transitions caused by
the phonon scattering~\cite{Ref30}. However, recent ARPES studies with 
soft X-rays have shown that this can be avoided and the band dispersion 
can be measured by cooling the sample~\cite{Ref31}. 
Therefore, the present ARPES 
experiments were performed at 50 K to minimize 
the influence of thermal diffuse scattering and hence maximize 
the direct transition component.

\section{Results and discussion}
In figs.~\ref{f.1}(f) and ~\ref{f.1}(g) we show the ARPES spectra of 
Ni(100) excited by 
780 eV photons with H- and V-polarization, from $\Gamma$ to $X$. The 
spectra show angular dependence in $\mathbf{k}$-space from $\Gamma$ 
to $X$ due to BDs, as marked with triangles. At the $\Gamma$-point 
in fig.~\ref{f.1}(f), the
 two peaks correspond to the critical points 
$\Gamma_{25'}$ and $\Gamma_{12}$ at 1.21 eV and 0.51 eV binding energy. 
Two bands disperse from $\Gamma_{25'}$ towards $X$-point and a third band 
disperses from $\Gamma_{12}$, corresponding to $\Delta_{2'}$, $\Delta_{5}$, 
and $\Delta_{2}$ bands, respectively (fig.~\ref{f.1}(f)). Features due to a 
fourth weak band ($\Delta_{1}$) are also barely seen. These BDs are 
almost consistent with ARUPS 
results~\cite{Ref10,Ref11,Ref12,Ref13,Ref14,Ref15}. Especially, the energy 
positions at the high symmetry $\Gamma$- and $X$-point are very consistent 
with (spin-integrated) ARUPS results indicated by bars in fig.~\ref{f.1}(f). 
Fig.~\ref{f.1}(g) (V-polarization)
reproduces the energy positions of the critical points $\Gamma_{12}$ 
and $X_{5\uparrow}$, with consistent BDs as in fig.~\ref{f.1}(f) 
(H-polarization), but enhances the $\Delta_{2}$ 
band relative to other bands~\cite{Ref32}. The results indicate that 
soft X-ray 
ARPES can be surely used to obtain BDs of solids. However, there is one 
important discrepancy in the present data compared with ARUPS results. 
The $\Delta_{2}$ band crosses the $E_{F}$ in the present spectra and its 
energy position at the $X$-point, that is $X_{2}$, is evidently above 
$E_{F}$. This is in contrast with ARUPS 
studies~\cite{Ref12,Ref14} which show that 
the majority and minority bands of $\Delta_{2}$ symmetry do not cross 
$E_{F}$, but remain below $E_{F}$ all along the $\Delta$-line.

\begin{figure}[b]
\onefigure[width=140mm]{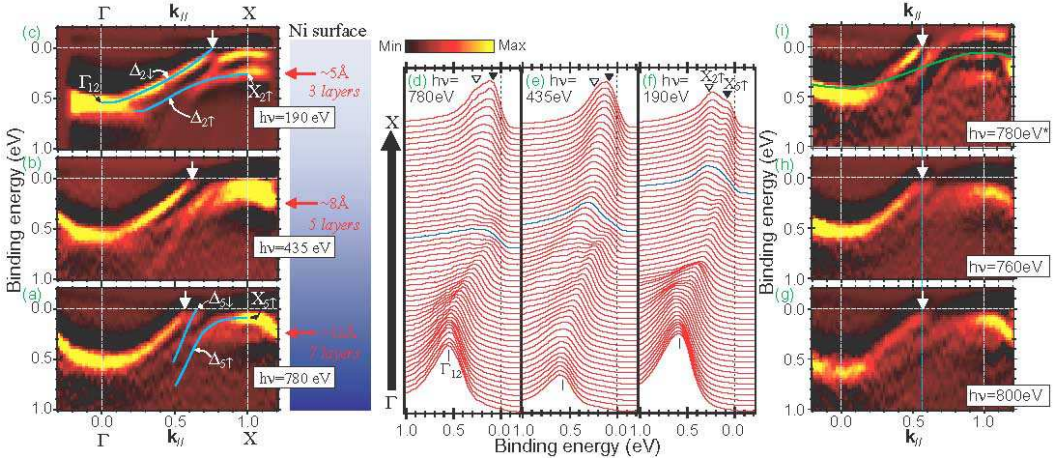}
\caption{Band maps of Ni(100) along $\Gamma$-$X$ 
($\Delta$-line) using (a) $h\nu$ = 780 eV, 
(b) 435 eV and (c) 190 eV, with V-polarization. Arrows indicate 
$\mathbf{k}$-points where the $\Delta_{2\downarrow}$ band crosses 
$E_{F}$. The blue lines show the observed band dispersions. 
The raw ARPES spectra with (d) 780 eV (e) 435 eV) and (f) 190 eV 
were used to obtain BMs of (a)-(c).
The spectra showing $E_{F}$ crossings are plotted with blue curves in (d)-(f). 
(g) and (h) show the BMs with 800 eV and 760 eV photons. 
Fig.\ 2(i) is obtained 
as I(780eV*) =
I(780) - 0.52I(435) 
(I: intensity weighted by atomic cross-sections after normalized by 
photon flux). The green line shows the $\Delta_{2\downarrow}$ 
band dispersion summarized from low-energy ARUPS studies and 
inverse photoemission studies \cite{Ref10,Ref11,Ref12,Ref13,
Ref14,Ref15,Ref16,Ref17,Ref18,Ref19}, which are reproduced by 
GWA + DMFT \cite{Ref25} calculation.}
\label{f.2}
\end{figure}

As a check of the $\Delta_{2}$ BD 
measured using $h\nu$ = 780 eV, and if truly so, to investigate its 
deviation from ARUPS results, we measured BDs with $h\nu$ = 435 
and 190 eV. In order to clearly see BDs, we plot band maps (BM)
(second derivative of raw spectra after smoothing). 
Figs.~\ref{f.2}(a)-(c) show BMs made using the raw spectra
shown in 
Figs.~\ref{f.2}(d)-(f)
obtained with V-polarization using 
$h\nu$ = 780, 435, 
and 190 eV, respectively. In fig.~\ref{f.2}(a), the $\Delta_{2}$ BD can be 
followed unambiguously and the $E_{F}$ crossing is clearly 
observed (arrow mark), corresponding to the $\Delta_{2\downarrow}$ 
minority band. 
The $\Delta_{2\uparrow}$ majority BD also can be seen in fig.~\ref{f.2}(a) 
but is more clear 
in figs.~\ref{f.2}(b) and ~\ref{f.2}(c), with $X_{2\uparrow}$ located at 
0.27 eV binding energy. 
The spectral intensity very close to the $E_{F}$ crossing of 
$\Delta_{2\downarrow}$ (fig.~\ref{f.2}(a), just next to the arrow mark), 
disperses to higher binding energy and is assigned to $\Delta_{5\downarrow}$. 
Additional confirmation comes from the 435 eV BM in fig.~\ref{f.2}(b) 
which shows 
nearly identical BDs, but for a small change of the $E_{F}$ crossing 
point in $\mathbf{k}$-space (arrow mark). The changed BD of the 
$\Delta_{2\downarrow}$ 
band with 435 eV photons (fig.~\ref{f.2}(b)) results in overlapping the 
$\Delta_{5\downarrow}$ band at $E_{F}$, which are separated with 
780 eV photons in fig.~\ref{f.2}(a). The most important result is 
obtained with the
$h\nu$ = 190 eV BM shown in fig.~\ref{f.2}(c). This BM shows 
well-resolved majority ($\Delta_{2\uparrow}$) and 
minority ($\Delta_{2\downarrow}$) bands. Probing just beyond the ARUPS 
regime changes BDs which are clearly intermediate to the bulk BDs 
(obtained with $h\nu$ = 435 and 780 eV) and surface BDs obtained by 
ARUPS~\cite{Ref10,Ref11,Ref12,Ref13,Ref14,Ref15}. 
The $\Delta_{2\downarrow}$ band crosses $E_{F}$ at 
$0.57$($\Gamma$-$X$) in fig.~\ref{f.2}(a), at $0.62$($\Gamma$-$X$) in 
fig.~\ref{f.2}(b), whereas it 
is $0.76$($\Gamma$-$X$) in fig.~\ref{f.2}(c). 
The shift of the $E_{F}$ crossing is also seen in the raw ARPES 
spectra (fig.~\ref{f.2}(d)-(f)), which show clear $E_{F}$ crossings 
with each photon energy(blue curves). The group velocity at $E_{F}$ 
also changes systematically from 0.62 eV\AA\ (190 eV), to 0.82 eV\AA\ 
(435 eV) and 1.11 eV\AA\ (780 eV). Since the 
$\Delta_{2\downarrow}$ BD changes systematically, the results indicate 
that the $\Delta_{2}$ bandwidth and the ($X_{2\uparrow} - X_{2\downarrow}$) 
ES is increased in the bulk compared to the ARUPS results. From ARUPS the 
ES is estimated to be 170 meV, an anomalously low ES measured only for the 
$X_2$ point on the surface~\cite{Ref12,Ref14,Ref19}. 

\begin{figure}[b]
\onefigure[width=80mm]{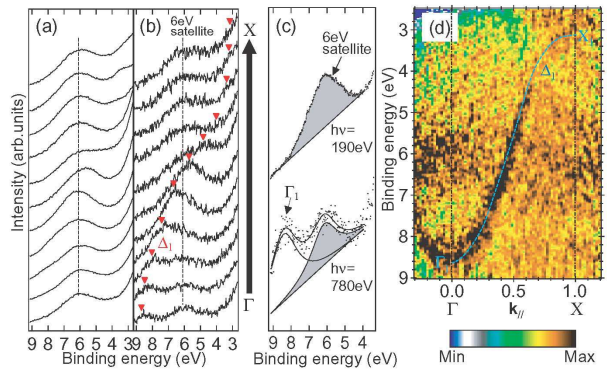}
\caption{ARPES spectra with $h\nu$ = (a)190 and (b)780 eV, for 
the 6 eV satellite from $\Gamma$ to $X$, normalized 
for area under the curve of the entire valence band. 
Triangles in (b) indicate $\Delta_{1}$ BD obtained from fig.\ 3(d). 
(c) A blow-up of 6 eV satellite feature at $\Gamma$-point for 
the same normalization. Curve fits (solid curves) indicate a reduction 
of $\sim$20$\pm$4\% in 6 eV satellite intensity for $h\nu$ = 780 eV
(shaded area), but no weakening of 
the correlation energy, consistent with its localized nature.  
(d) The BM between 2.5 and 9.0 eV binding energy 
showing $\Delta_{1}$ BD (broken curve).}
\label{f.3}
\end{figure}

We next check how the $E_{F}$ crossing of $\Delta_{2\downarrow}$ is 
sensitive to the $\mathbf{k}_{\bot}$-point in Ni band structure. The 
BMs with 800 eV and 760 eV photons (fig.~\ref{f.2}(g) and (h)), 
which corresponds to $\approx \pm$10 \% variation in 
$\mathbf{k}_{\bot}$ along $\Gamma$-$X$(see fig.~\ref{f.1}(a))
as compared to data obtained with 780 eV photons (fig.~\ref{f.2}(a)) 
show very similar BDs and also the $E_{F}$ crossing 
of $\Delta_{2\downarrow}$ as in fig.~\ref{f.2}(a). Therefore, the BD 
is insensitive to the $\mathbf{k}_{\bot}$ variations in the range 
of $\sim$ 0.1($\Gamma$-$X$) for
high energies and the change in $E_{F}$ crossing is not due to a change 
of the $\mathbf{k}_{\bot}$-point.

Since, at any photon energy, the experiments actually measure an 
integral of the intensity, weighted by an exponential factor for 
the MFP, we try to extract the bulk BD from ARPES with 780 eV 
photons. The spectra with 780 eV photons is estimated to roughly 
contain about 50\% contribution from the top 4-5 layers, which is 
the probing depth at $h\nu$ = 435 eV. 
The ARPES spectra with 780 eV photons is 
subtracted by the 435 eV spectra weighted 
by the exponential factor (0.52) and assuming atomic cross-sections. 
The result is shown in fig.~\ref{f.2}(i) and confirms that the BD seen 
for bulk Ni are similar to the data of fig.~\ref{f.2}(a). In particular, 
the separate BDs are more clear at the $X$ point. The data thus 
prove that the $\Delta_{2\downarrow}$ BD is different in the bulk 
compared to surface sensitive ARUPS studies(overlaid as green line).

Thus, the reduced probing depth measurements indicate bulk in-plane BD of 
$\Delta_{2\downarrow}$ band systematically connects to the ARUPS 
results~\cite{Ref10,Ref11,Ref12,Ref13,Ref14,Ref15}. 
The $\Delta_{2\downarrow}$ band which is pinned just 
below $E_{F}$ at $X$-point ($X_{2\downarrow}$) in surface sensitive 
ARUPS, crosses $E_{F}$ in bulk sensitive ARPES with $X_{2\downarrow}$ 
lying above $E_{F}$. This leads to an additional minority spin 
$X_{2\downarrow}$ hole pocket. There exists 
an early spin polarized study by Kisker et al.~\cite{Ref33} showing 
that the $X_{2\downarrow}$ state lies above $E_{F}$ although later studies 
have concluded it lies below $E_{F}$. In view of the present results, 
the larger MFP used but at very low energy ( 4-6 eV ) by Kisker 
et al.~\cite{Ref33} reflects the bulk BS~\cite{Ref04}. While de 
Haas-van Alphen measurements also showed only the $X_{5\downarrow}$ hole 
pocket along $\Gamma$ to $X$~\cite{Ref34}, magneto-crystalline anisotropy 
measurements~\cite{Ref35} indicated evidence for the 
additional $X_{2\downarrow}$ hole pocket.

The layer dependent BD is coupled to the widening of the $\Delta_{2}$ 
band width observed in ARPES using high $h\nu$, suggesting weaker 
electron-electron correlations in the bulk. Theoretical 
calculations which include 
Coulomb correlations show that $X_{2\downarrow}$ is located below $E_{F}$
~\cite{Ref14,Ref19,Ref23,Ref24,Ref25} in contrast to s-LDA 
calculations~\cite{Ref22} and the 6 eV satellite is also 
reproduced~\cite{Ref23,Ref24,Ref25}. 
Hence, we checked for variations 
of ARPES spectra in the 6 eV satellite region as a function of $h\nu$ 
(figs.~\ref{f.3}(a) and ~\ref{f.3}(b)). The ARPES spectra show the 
$\Delta_{1}$ band 
dispersing across and overlapping the 6 eV satellite between $\Gamma$ 
to $X$, as expected from s-LDA calculations~\cite{Ref22} and ARUPS 
studies~\cite{Ref10,Ref11,Ref12,Ref13,Ref14,Ref15}. Even though 
the $4sp$ character $\Delta_{1}$ band 
crossing makes quantification very 
difficult, an attempt to do so with the $\Gamma$ point spectra 
(fig.~\ref{f.3}(c)), 
where the $\Delta_{1}$ band is separated, shows a reduction 
of $\sim$20$\pm$4\% in the 6 eV satellite intensity on increasing $h\nu$ from 
190 to 780 eV. This reduction in intensity is small and not conclusive 
but suggestive of reduced correlations 
in the bulk~\cite{Ref16,Ref17,Ref18,Ref23}, consistent with the 
layer dependent change of $\Delta_{2\downarrow}$ band dispersion in 
fig.~\ref{f.2}. Also, since the electron correlation in Nickel leads 
to a reduced ES compared to s-LDA calculation~\cite{Ref19,Ref25}, 
the reduced 
correlation is consistent with the enhancement of ES observed at $X_{2}$. 
Since the 6 eV satellite feature is still observed with 780 eV photons as in 
angle-integrated X-PES studies~\cite{Ref16}, it shows that 
ARPES is necessary to check for the bulk to surface changes in 
the valence BS. In fig.~\ref{f.3}(d), 
we plot a BM for 2.5 to 9.0 eV binding energy and $h\nu$\ = 780 eV, 
where $\Delta_{1}$ BD is clearly seen, consistent with 
ARUPS results. We have also confirmed that the $\Delta_{1}$ 
BD is similar to that obtained by applying the 
sum rule as discussed in ref.~\cite{Ref36}.

Another important point to note is that the $\Delta_{5}$ and $\Delta_{2'}$ 
(fig.~\ref{f.1}(f)) BDs do not change with $h\nu$ i.e.\ as a function 
of probing depth. Since $\Delta_{5}$ and $\Delta_{2'}$ band originate 
in $t_{2g}$-type states, it is not modified by electron correlations 
like the $e_{g}$-derived states~\cite{Ref23}. 
Even between the 
$e_{g}$-derived states, the $\Delta_{2}$ band ($d_{x^{2}-y^{2}}$) possibly 
shows more layer dependent correlation than the $\Delta_{1}$ band 
($d_{z^{2}-r^{2}}$) since the ($d_{x^{2}-y^{2}}$) orbital extends 
outwards along the surface normal direction (100). This suggests similarity 
to recent results showing varying correlations 
between the $t_{2g}$-orbitals of ($d_{xy}$), ($d_{xz}$), and ($d_{yz}$) 
symmetry for correlated oxides~\cite{Ref05}.

\section{Conclusion}
In conclusion, tunable soft X-ray polarization dependent ARPES 
makes it possible to resolve the changes in BDs with specified 
symmetry and layers. We observe layer dependent 
$\Delta_{2\downarrow}$ band dispersion, which leads to an additional minority 
spin $X_{2\downarrow}$ hole pocket. However, the $\Delta_{5}$\ and 
$\Delta_{2'}$\ 
($t_{2g}$)-symmetry bands are not affected by the existence 
of the surface. Thus, the observed bulk versus surface difference in 
behavior of the electronic states depends on the band symmetry. 
The results show layer dependent BD changes, suggesting 
correlation effects and ES being coupled to changes in the probing 
depth, and the importance of Coulomb correlations in theoretical 
calculations for the surface electronic BS. 
Soft X-ray ARPES is thus shown to be a very important tool 
to study layer dependent BD of 3-D solids.

\acknowledgments
We thank Drs.\ H. Ohashi, Y. Tamenori, T. Ito, and P. A. Rayjada 
for help with experiments and Profs.\ A. Fujimori, 
T. Yokoya, A. Kotani, M. Taguchi, M. Usuda, T. Jo, and 
Dr.\ K. Horiba for valuable discussions.


\begin{thebibliography}{0}
\bibitem{Ref29}
  This is understood from the measured spectra which show constant 
  shifts in $\mathbf{k}_{\|}$ by $0.22(\Gamma - X)$, while the 
  $\mathbf{k}_{\bot}$ negligibly shifts, for 780 eV. Similarly, for 
  $h\nu$ = 435 eV and 190 eV, constant $\mathbf{k}_{\|}$ shifts of 
  $0.12(\Gamma - X)$ and $0.05(\Gamma - X)$ were determined. The shift 
  of $\mathbf{k}$ is identified from accurate determination of the high 
  symmetry points $\Gamma$ and $X$.
\bibitem{Ref32}
  In fig.~\ref{f.1} (d) and (e), $\Delta_{1}$ band should 
be mainly observed due to dipole selection rules and the intensity 
decrease at $X$-point seen in the figure is 
caused by the $\Delta_{1}$ band moving above $E_{F}$, as also expected 
from ARUPS. In fig.~\ref{f.1} (f) and (g), on the other hand, bands with all 
symmetries along $\Delta$ can be observed because of the measurement 
of the in-plane band dispersion, and the relative intensity of each band 
is governed by matrix elements. The differences in spectra 
is also attributed to the different volumes sampled with the different 
photon energies used.
\end{thebibliography}
\end{document}